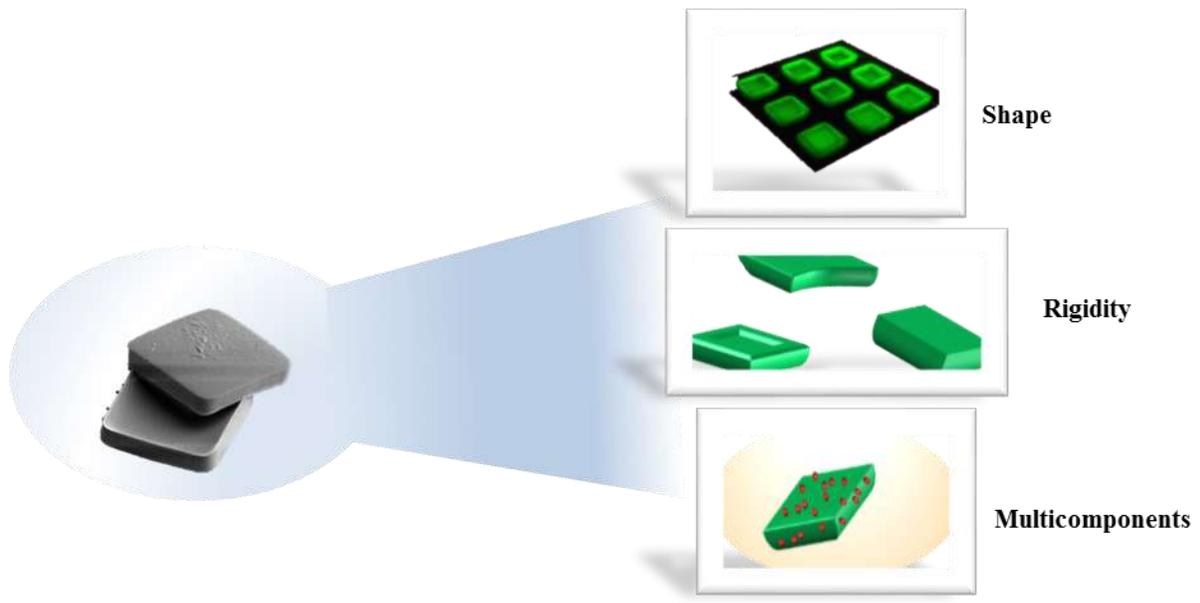

**GRAPHICAL ABSTRACT**

# HIERARCHICAL MICROPLATES AS DRUG DEPOTS WITH CONTROLLED GEOMETRY, RIGIDITY AND THERAPEUTIC EFFICACY

Martina Di Francesco[1], Rosita Primavera[1], Davide Romanelli[1], Roberto Palomba[1], Tiziano Catelani[1], Christian Celia[2], Luisa Di Marzio[2], Massimo Fresta[3], Daniele Di Mascolo[1], Paolo Decuzzi[1,*]

[1] Laboratory of Nanotechnology for Precision Medicine,

Fondazione Istituto Italiano di Tecnologia,

Via Morego 30, Genoa 16163, Italy

[2] Department of Pharmacy, University of Chieti – Pescara "G. D'Annunzio",

Via dei Vestini - Campus universitario - 66100 Chieti, Italy

[3] Department of Health Sciences,

University of Catanzaro "Magna Graecia", Viale Europa - 88100 Catanzaro, Italy

[*] To whom correspondence should be addressed: Paolo Decuzzi, PhD. Phone: +39 010 71781 941, Fax: +39 010 71781 228, E-mail: Paolo.Decuzzi@iit.it

**ABSTRACT**

A variety of microparticles have been proposed for the sustained and localized delivery of drugs whit the objective of increasing therapeutic indexes by circumventing filtering organs and biological barriers. Yet, the geometrical, mechanical and therapeutic properties of such microparticles cannot be simultaneously and independently tailored during the fabrication process in order to optimize their performance. In this work, a top-down approach is employed to realize micron-sized polymeric particles, called microPlates (μPLs), for the sustained release of therapeutic agents. μPLs are square hydrogel particles, with an edge length of 20 μm and a height of 5 μm, made out of poly(lactic-co-glycolic acid) (PLGA). During the synthesis process, the μPL Young's modulus can be varied from 0.6 to 5 MPa by changing PLGA amounts from 1 to 7.5 mg, without affecting the μPL geometry. Within the porous μPL matrix, different classes of therapeutic payloads can be incorporated including molecular agents, such as the anti-inflammatory dexamethasone (DEX), and nanoparticles, containing themselves imaging and therapeutic molecules. As a proof of principle, μPLs are loaded with free DEX and 200 nm spherical polymeric nanoparticles, carrying DEX molecules (DEX-SPNs). Electron and fluorescent confocal microscopy analyses document the uniform distribution and stability of molecular and nano-agents within the μPL matrice. This multiscale, hierarchical microparticle releases DEX for at least 10 days. The inclusion of DEX-SPNs serves to minimize the initial burst release and modulate the diffusion of DEX molecules out of the μPL matrix. The pharmacological and therapeutic properties together with the fine tuning of geometry and mechanical stiffness make μPLs a unique polymeric depot for the potential treatment of cancer, cardiovascular and chronic, inflammatory diseases.

**INTRODUCTION**

The systemic administration of nanoparticles for the detection and treatment of diseased tissues is progressively shaping up as a successful strategy, as demonstrated by the growing number of clinical trials based on nanomedicines.[1] The biophysical mechanisms regulating the accumulation of blood-borne nanoparticles within malignant tissues are mostly associated with the tortuosity and hyper-permeability of the diseased microcirculation, as compared to normal vascular beds.[2-3] For the same mechanisms, a portion of intravenously injected nanoparticles tends also to accumulate within filtering organs, such as the liver and spleen, which are characterized by discontinuous and permeable vascular walls.[4] A variety of approaches have been proposed to reduce the unspecific deposition of nanomedicines within these filtering organs, including the proper tailoring of the nanoparticle surface, geometrical and mechanical properties.[5-7] An alternative approach relies on drug depots, which follow a different strategy. These are realized by the intra-tissue deposition of nano- and micro-particles, nanofibers and gels, and allow for the sustained and long-term release of large amounts of therapeutic agents. Importantly, these agents are directly deployed from the depot within the diseased tissue bypassing filtering organs and biological barriers, such as the blood brain, the intestinal and lung epithelial barriers.[8] As such, drug depots could represent a valuable alternative to the systemic administration of therapeutics in a number of applications. This has been already shown clinically and pre-clinically for cancer, cardiovascular, and localized, chronic inflammatory diseases.[9-11]

In the post-surgical treatment of brain cancers, the local delivery of carmustine from biodegradable polymeric wafers is commonly practiced in the clinic and has contributed to improve the life expectation of patients diagnosed with high grade gliomas.[9, 12] In this application, polymeric wafers are deposited on the surface of the tumor's resected cavity and release their cytotoxic content for about 6 days with the objective of removing residual cancer cells. A similar strategy has been proposed for the treatment of other deadly malignancies, including pancreatic, liver and lung carcinomas.[13-15] Drug depots have been also developed for the management of cardiovascular diseases, chronic inflammatory diseases, and brain disorders. For instance, drug-coated balloons and medicated stents are used for modulating the risk of restenosis in atherosclerosis.[16-17] After recanalization, thin films of cytotoxic agents are deposited onto the injured vasculature with the objective of inhibiting local cell growth and the risk of *de novo* occlusions. In osteoarthritis, nano and microparticles of chitosan and other polymers have been loaded with a variety of agents, including anti-inflammatory molecules, analgesics, growth factors, and injected intra-articularly to induce tissue repair and resolve local inflammation.[18-19] Similar approaches have been proposed for endodontic applications where nano and microparticles were used for the local delivery of anti-microbial and anti-inflammatory agents, and growth factors.[20-21] Finally, micron-sized polymeric particles have been proposed for delivering drugs directly within specific areas of the brain for the treatment of diverse disorders, including epilepsy[22], Parkinson's disease[23], and behavioral disorders.[24]

As described above, most local drug depots are realized with nano/micro-particles, nanofibers, polymeric films, and gels directly inoculated at the diseased site.[8] The efficacy of the system is dictated by the loading and release profiles of the active agents as well as by the spatial distribution and integration of the depot with the surrounding tissue.[8, 25] Following this notion, in this work, innovative modular and hierarchical polymeric microparticles – the microplates (µPLs) – are realized using a top-down fabrication approach where the size, shape, surface properties and mechanical stiffness can be simultaneously and independently tailored during the synthesis process.[26-27] In the present configuration, µPLs present a square base of 20 µm, a height of 5 µm, are made out of PLGA and their stiffness is finely tuned between about 0.5 and 5 MPa. Furthermore, µPLs are loaded with different molecular agents and nanoparticles returning a multifunctional, hierarchical system for local drug delivery. As a proof of principle, µPLs are loaded with the anti-inflammatory molecule dexamethasone acetate (DEX) and with 200 nm spherical nanoparticles, loaded themselves with DEX. The anti-inflammatory efficacy of µPLs is tested on two phagocytic cell types, namely RAW 264.7 cells and bone marrow derived monocytes (BMDMs).

## MATERIALS AND METHODS

### MATERIALS

Polydimethylsiloxane (PDMS) (Sylgard 184) was purchased from Dow Corning (Midland, Michigan, USA). Poly(vinyl alcohol) (PVA), poly(lactic-co-glycolic) acid (PLGA) (50:50), dexamethasone acetate (DEX), rhodamine B (RhB), MTT assay, sodium phosphate dibasic dehydrate, sodium phosphate monobasic monohydrate, acetonitrile, macrophage colony-stimulating factor (M-CSF), bacterial lipopolysaccharides (LPS) paraformaldehyde (PFA) and trifluoroacetic acid (TFA) were purchased from Sigma Aldrich (Saint Louis, Missouri, USA). All the lipids were purchased from Avanti Polar Lipid: 1,2-distearoyl-sn-glycero-3-phosphoethanolamine-N-[carboxy(polyethylene glycol)-2000] (sodium salt) (DSPE-PEG-COOH) and 1,2-dipalmitoyl-sn-glycero-3-phosphocholine (DPPC). Curcumin (CURC) was purchased from Alfa Aesar (Haverhill, Massachusetts, USA). 20 nm gold nanoparticles were purchased from Nanopartz Inc. (USA) Polycarbonate membrane filters were purchased from Sterlitech Corporation (USA). Wheat germ agglutinin (WGA), 4',6-diamidino-2-phenylindole (DAPI), RNAeasy Plus Mini Kit (50) purchased from Qiagen. PBS, RT-PCR reagents and gene specific primers were purchased from Thermo Scientific (USA). RAW 264.7 cell line was obtained from American Type Culture Collection (ATCC, LGC Standards, Teddington, UK). High-glucose Dulbecco's modified Eagle's Minimal Essential Medium (DMEM) and heat-inactivated fetal bovine serum (FBS) were obtained from GIBCO (Invitrogen Corporation, Giuliano Milanese, Milan, Italy). Lowicryl® K11M resin was purchased from Chemische Werke Lowi (Germany).

**METHODS**

**Fabrication Process of the MicroPlates.** The synthesis of the microplates (µPLs) follows a multi-steps process, partially described in previous works by the authors and other scientists.[28-29] First, a silicon master template is fabricated using Direct Laser Writing (DLW). The master template is a silicon substrate on which arrays of square wells are realized with an edge length of 20 µm and a depth of 5 µm, and each arrays is separated by a 3 µm gap. Then, a PDMS replica of the silicon master template is obtained by covering it with a mixture of PDMS and elastomer (10:1, v/v). The resulting sample is left in a vacuum chamber to remove bubbles formed during the mixing process and polymerized at 60 °C for 4h. Then, the PDMS template is peeled off the silicon substrate and used to obtain a PVA template, by pouring a PVA solution (3.5% w/v, in deionized water) on its patterned surface. The resulting PVA film is dried at 60 °C and finally peeled off the PDMS template. Thus, the PVA film has the same arrays of wells as the original silicon master template. In the last step, µPLs are obtained by spreading the mixture of PLGA and the desired payload onto the PVA wells. After solvent evaporation, the loaded PVA templates are dissolved in DI water at room temperature in ultrasonic bath. Released µPLs are purified from PVA solution by using polycarbonate membrane filters (30 µm) and collected through sequential centrifugations (1,717 *g* for 5 min). Different amounts of PLGA, namely 1, 5 and 7.5 mg, are used in order to modulate µPL rigidity. µPLs loaded with dexamethasone acetate (DEX-µPLs) or curcumin (CURC-µPLs) are realized by dispersing within the polymeric mixture 500 µg of dexamethasone acetate (DEX) or curcumin (CURC), respectively. For the degradation studies, 2 µg of fluorescent Rhodamine B (RhB) are dissolved in the polymeric mixture, thus replacing the drug molecules.

**Physicochemical characterization of the MicroPlates.** The geometry and physicochemical properties of µPLs are characterized using different techniques. µPLs size and shape analyses are performed via scanning electron microscopy (SEM, Elios Nanolab 650, FEI). Briefly, a drop of sample is spotted on a silicon template and uniformly sputtered with gold to increase the contrast and reduce sample damaging. An acceleration voltage of 5 - 15 keV is used for SEM imaging. µPLs average size and distribution are also obtained using a Multisizer 4 Coulter Particle Counter (Beckman Coulter, CA). Briefly, µPLs are resuspended in the electrolyte solution and analyzed, following protocols described by the vendor. The µPLs electrostatic surface charge is determined using a Zetasizer Nano (Malvern, U.K.).

**Biofrmaceutical characterization of the MicroPlates.** For measuring DEX loading and encapsulation efficiency (LE and EE, respectively) into DEX-µPLs, samples are lyophilized, dissolved in acetonitrile/$H_2O$ (1:1, v/v) and analyzed by high-performance liquid chromatography (HPLC) (Agilent 1260 Infinity, Germany), equipped with a 100 µl sample loop injector. A C18column (4.6×250 mm, 5 µm particle size, Agilent, USA) was used for the chromatographic separation. The DEX was eluted under isocratic condition using a binary solvent system (H$_2$O+0.1% (v/v) TFA: AcN+0.1% (v/v) TFA, 50:50 v/v) pumped at a flow rate of 1.0 ml/min. The UV detection was set at 240 nm. The retention time of DEX was 3.65 min. An external standard curve in a linear concentration ranging from 1 to 300 µg/ml was used for the quantification of DEX. A standard solution (10 mg/ml in acetonitrile) was used for the construction of the standard curve. The amount of DEX was determined using the following equation (r2= 0.9997):

Eq. 1 $\qquad$ AUC = 2.5118x - 1.338

where x is the drug concentration (µg/ml). HPLC analysis showed that no interference was determined by the various components of MicroPlates.

LE and EE are determined using the following equations:

$$LE(\%) = \frac{DEX\ weight\ in\ particles}{Total\ weight\ of\ the\ particles} \times 100$$

$$EE(\%) = \frac{DEX\ weight\ in\ particles}{DEX\ initial\ feeding\ amount} \times 100$$

To study DEX release kinetics, 200 µl of DEX-µPLs solution are placed into Slide-A-Lyzer MINI dialysis microtubes with a molecular cut-off of 10 kDa (Thermo Scientific) and then dialyzed against 4 L of PBS buffer (pH 7.4) or phosphate saline buffer (pH 5.5, 1 X) at 37 °C. For each time point, three samples are collected and centrifuged (1,717 *g* for 5 min). Pellets are then dissolved in acetonitrile/H$_2$O (1:1, v/v) and analyzed by HPLC. The experimental data are fitted by using the Ritger-Peppas model for controlled not swellable drug delivery systems:[30]

$$Y = K^* x^n$$

where, Y represents the drug percentage released, x is the time of observation, and *k* and *n* are the fitting parameters.

**Stability of the MicroPlates.** The stability of µPLs is evaluated by fluorescence microscopy (Leica 6000, Wetzlar, Germania) and SEM analysis. 200 µl of empty and RhB-µPLs were incubated in PBS buffer (pH 7.4) or phosphate saline buffer (pH 5.5, 1 X) under mechanically stirring at 37 °C. At different time points (1, 3, 6 and 10 days), samples are analyzed to monitor structural and morphological changes.

**Toxicity and therapeutic efficacy of the dexamethasone-loaded MicroPlates.** RAW 264.7 cells are cultured at 37 °C in 5% $CO_2$, in high-glucose DMEM, supplemented with 10% FBS and 1% L-Glutamine, according to ATCC instructions. Cells are seeded into 96-well plates at a density of $5 \times 10^3$ cells per well and incubated for 24h. Cells are treated with different concentrations of free DEX and DEX-µPLs (namely, 0.1, 0.5, 1, 10 and 30 µM of DEX in both cases) or empty µPLs, at cell:µPLs ratios of 1:0.01, 1:0.05, 1:0.1, 1:1 and 1:4, to match the number of µPLs used in the case of DEX-µPLs. MTT solution is added for 4 h and formazan crystals dissolved in ethanol. Absorbance is measured at 570 nm, using 650 nm as reference wavelength (Tecan, Männedorf, Swiss). The percentage of cell viability is assessed according to the following equation:

$$Cell\ viability\ (\%) = \frac{AbsT}{AbsC} x 100$$

where AbsT is the absorbance of treated cells and AbsC the absorbance of control (untreated) cells.

To investigate the anti-inflammatory activity of DEX-µPLs towards stimulated macrophages, the expression levels of three pro-inflammatory cytokines, namely TNF-α, IL-1β, and IL-6, are evaluated in RAW 264.7 cells and rat bone marrow-derived macrophages (BMDMs). For BMDMs harvesting, rat femurs are flushed 4 times with 500 µl of PBS after cutting bones extremities. Cells suspension is filtered using 70 µm cell strainers and centrifuged at 72 $g$ for 8 min. Cells are plated in Petri dishes and after 3 days, the medium is changed in order to remove unattached cells. BMDMs are used after 4 more days. BMDMs are cultured in DMEM supplemented with 15% FBS, 1% penicillin/streptomycin and rat M-CSF (according to vendor indications). Both cells are cultured in controlled environmental conditions (37 °C in 5% $CO_2$) and are seeded into 6-wells plates at a density of $3\times10^5$ cells per well for 24h.

Cells are treated with DEX-µPLs at different concentrations and incubated for 5 h. Then, bacterial lipopolysaccharides (LPS) are added at a concentration of 100 ng/ml and incubated for 4 h. At the end, RNA is extracted using RNAeasy Plus Mini Kit (Qiagen) and quantified by NanoDrop2000 (Thermo Scientific, Waltham, Massachusetts, USA). Real-time RT-PCR is used to measure mRNA levels of inflammatory cytokines. For each condition, samples are in triplicate. RT-PCR reactions are carried out using a Power SYBR Green RNA-to-CT 1-Step Kit (Applied Biosystems). Reactions are performed in a final volume of 20 µl. Oligonucleotide primer pairs are as follows: for GAPDH, 5'-GAACATCATCCCTGCATCCA-3' and 5'-CCAGTGAGCTTCCCGTTCA-3'; for TNF-α, 5'-GGTGCCTATGTCTCAGCCTCTT-3' and 5'-GCCATAGAACTGATGAGAGGGAG-3'; for IL-1β, 5'-TGGACCTTCCAGGATGAGGACA-3' and 5'-GTTCATCTCGGAGCCTGTAGTG-3'; for IL-6, 5'-TACCACTTCACAAGTCGGAGGC-3' and 5'-CTGCAAGTGCATCATCGTTGTTC-3'.

**Mechanical characterization of the MicroPlates.** The atomic force microscopy (AFM) is used to measure the stiffness of µPLs. Before the deposition of the sample, to promote adhesion and prevent µPLs movement during the analysis, few drops of $10^{-2}$ M poly(iminoethylene) (PEI) are spotted on a microscope slide, since PEI is positively charged and supports electrostatic interactions with the negatively charged PLGA carboxylic terminations. After few minutes, the excess of PEI is removed and 10 µl of sample are placed on the slide. AFM analysis is performed using the Nanowizard II AFM (JPK Instruments, Berlin, Germany), mounted on an Axio Observer D1 inverted optical microscope (Carl Zeiss, Oberkochen, Germany). Quantitative imaging (QI) data set are acquired in liquid, to better mimic physiological environment, using DNP V-shaped silicon nitride cantilevers, with nominal spring constant ranging from 0.12 to 0.48 N/m, resonance frequency in air 40-75 kHz, and a silicon nitride tip with typical curvature radius 20-60 nm (Bruker, Billerica, MA, US). The maximum force applied to the sample is 1 nN.

**Preparation of Spherical Polymeric Nanoparticles (SPNs).** Spherical polymeric nanoparticles are prepared by a slightly modified oil-in-water (O/W) emulsion/solvent evaporation technique, as compared to previous work of the authors and other scientists.[31-32] Briefly, 300 µl of DEX solution (6.66 mg/ml, in acetone) is added at 450 µl of organic solution containing PLGA (1 mg, in chloroform), DPPC (100 µg in chloroform) and mixed using a probe sonicator for 15 sec at 60% Amplitude. This mixture is added drop by drop into 3 ml of 4% ethanol containing 110 µg of DSPE-PEG-COOH and sonicated for 1 min at 60% Amplitude. The resulting emulsion is placed under reduced pressure and magnetic stirring to foster the evaporation of the organic solvents. After the complete removal of organic solvents, the obtained nanoparticles are purified through centrifugation. To remove the debris of the synthesis procedure, at first the solution is centrifuged at 452 $g$ for 2 min, and then SPNs undergo several centrifugations at 18,213 $g$ for 15 min to remove the unloaded drug. In order to demonstrate SPNs integrity after loading into µPLs, DEX is replaced with gold nanoparticles, and part of DSPE-PEG is substituted with DSPE-Cy5.

**Characterization of DEX-loaded Spherical Polymeric Nanoconstructs.** Average size, size distribution and zeta potential of DEX-SPNs are analyzed using Dynamic Light Scattering (DLS). Samples are diluted with isosmotic double distilled water (1:10 v/v) to avoid multiscattering phenomena and analyzed at 25 °C with Zetasizer Nano (Malvern, U.K.), , equipped with a 4.5 mW laser diode, operating at 670 nm as a light source, and the scattered photons detected at 173°. A third order cumulative fitting autocorrelation function was applied to measure average size and size distributions. The analysis was carried out according to the following instrumental set up: (a) a real refractive index of 1.59;(b) an imaginary refractive index of 0.0; (c) a medium refractive index of 1.330; (d) a medium viscosity of 1.0 mPa × s; and (e) a medium dielectric constant of 80.4. DEX loading (LE) and encapsulation efficiency (EE) into DEX-SPNs are measured as previously reported. For the release kinetics, 200 µl of samples were placed into Slide-A-Lyzer MINI dialysis microtubes with a molecular cut-off of 10 kDa (Thermo Scientific) and then dialyzed against 4 L of PBS buffer (pH 7.4) at 37 °C. For each time point, three samples were collected and centrifuged at 18,213 $g$ for 15 min. Pellets were then dissolved in acetonitrile/$H_2O$ (1:1, v/v) and analyzed by HPLC.

**Preparation and characterization of SPNs-loaded MicroPlates.** DEX-SPNs-µPLs are synthesized using the protocol used for DEX-µPLs fabrication, with some modifications. To prevent DEX-SPNs damaging during the synthesis process, they are added to a 20% (w/v) PVA aqueous solution, under magnetic stirring, for 15 min, in order to get a PVA coating. Subsequently, PVA coated DEX-SPNs are purified using centrifugation and finally added to the polymeric mixture (5 mg of PLGA) and spread onto the PVA template. Then DEX-SPNs-µPLs are released, purified, and characterized as above reported for DEX-µPLs.

To assess their distribution and their integrity after the loading process into µPLs, gold nanoparticles-loaded Cy5-tagged-SPNs (Au-Cy5-SPNs) are loaded into CURC-µPLs and then analyzed using Nikon A1 confocal microscopy (Dexter, MI) and a JEOL JEM 1011TEM , operating with an acceleration voltage of 100 kV and equipped with a 11 Mp fiber optical charge coupled device (CCD) camera (GATAN Orius 830). For TEM analysis, µPLs from a suspension are dried onto glass slide and embedded into Lowicryl® K11M resin. The resin is polymerized for 48h under UV lamp (wavelength 365 nm) and then cut using LEICA EM UC6 ultra-microtome and ultra sonic 35° DIATOME diamond knife.

Finally, for determine any possible interaction with other cells, more similar to soft tissues, human fibroblasts cells were cultured at 37 °C in 5% $CO_2$, in DMEM, supplemented with 10% FBS and 1% L-Glutamine. Cells were seeded into 8 chambered cover glass system (Lab Teck II, Thermo Scientific, USA) at a density of $20 \times 10^3$ cells per well and incubated for 24h. Cells are incubated with Au-Cy5-SPNs-loaded CURC-µPLs for 1h. Cells are fixed using 4% PFA and stained with WGA and DAPI, according to vendor indications. Samples were analyzed using confocal microscopy Nikon A1 (Dexter, MI).

**Statistics analysis**. All data are represented as the average ± standard deviation (SD) of 3 different measurements, unless differently specified. The statistical significant difference was assessed using ANOVA test, with Tukey's Multiple Comparison Test as post-hoc test. For the fitting of release curves, the extra sum-of-square F test is performed. A p value ≤ 0.05 is considered statistically significant.

**RESULTS AND DISCUSSION**

**Preparation and physicochemical characterization of MicroPlates.** Microplates (µPLs) were prepared following different sequential steps, as schematically reported in **Figure.1A**. The first step was the fabrication of a silicon master template with specific geometrical features. This original template was replicated into a polydimethylsiloxane (PDMS) template and then into a sacrificial polyvinyl alcohol (PVA) template, as described in the Methods section. A polymeric mixture was carefully deposited on the PVA template to fill up all square wells. Finally, the PVA template loaded with the polymeric mixture was dissolved in water, thus releasing the µPLs. The pictures in **Figure.1B** show, from left to right, a scanning electron microscopy (SEM) image of the original silicon master template with arrays of wells; a SEM microscopy image of the PDMS template with arrays of pillars; a fluorescent microscopy image of a PVA template carrying curcumin-loaded µPLs (green fluorescence); and a fluorescent microscopy image of free curcumin-loaded µPLs, released from the PVA template.

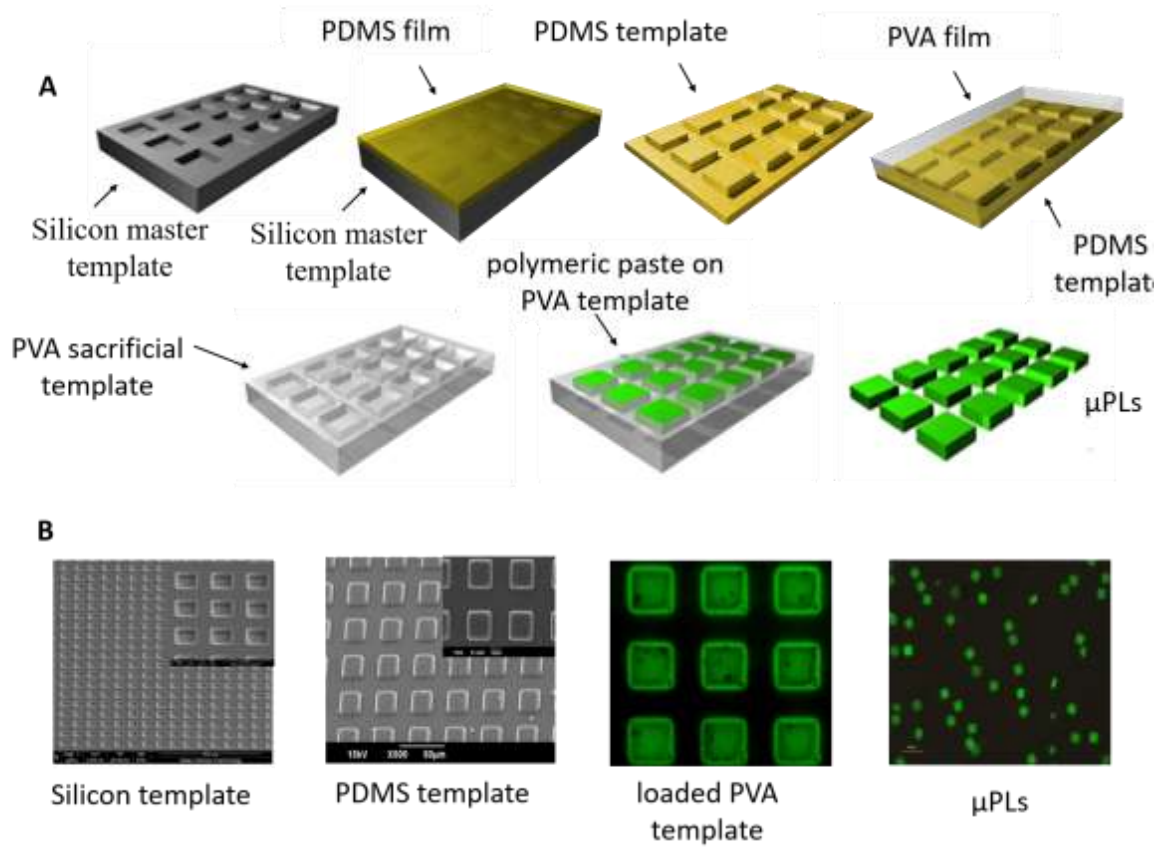

**Figure.1. MicroPlates (µPLs) fabrication process. A**. Sequential steps in the fabrication of 20 × 20 × 5 µm MicroPlates (µPLs). A silicon master template (grey) is fabricated via direct laser writing and replicated into a PDMS template (yellow), whose pattern is then transferred into a sacrificial PVA template (white). This PVA template is loaded with the polymeric paste constituting the final µPLs and enclosing the imaging and therapeutic payloads. µPLs are released and collected upon dissolution in DI water of the sacrificial PVA template; **B**. From the left to the right, SEM images of silicon and PDMS templates, and optical microscopy images of the PVA loaded with the polymeric paste and released µPLs.

Curcumin for its intrinsic fluorescence was used for microscopy analysis. Curcumin-loaded µPLs (CURC- µPLs) were obtained by depositing a paste of poly(lactic-co-glycolic acid) (PLGA) and curcumin within the sacrificial PVA template. MicroPlates were characterized using different techniques (**Figure.2**). In particular, a scanning electron microscopy (SEM) image is reported in **Figure.2A**, showing the specific geometry of µPLs: a square shape with edge length of about 20 µm and a height of about 5 µm, which precisely fit the size and shape of the wells in the original master silicon template. A size distribution analysis via Multisizer showed a single peak of ~15 µm with a relatively narrow distribution (**Figure.2B**). Indeed, given the non-spherical shape of the µPLs, the instrument returns an average characteristic size rather than the actual edge length of the particle. The µPL eletrostatic surface charge was of -17.9 ± 5.1 mV, which is due to the carboxylic termination on the PLGA chains. **Figure.2C** shows a confocal fluorescent image of CURC-µPLs inside the PVA template as maximum intensity projections. Side projections of a single µPL are also provided, demonstrating the homogeneous distribution of curcumin within the PLGA matrix and confirming the actual square geometry of the µPLs. **Figure.2D** shows a tri-dimensional reconstruction of a single µPL. Altogether the data suggest that this top-down fabrication strategy can precisely tailor the size and shape of µPLs, as already shown by the authors at smaller scales.[26] Moreover, as reported in **Figure.2E**, µPLs are comparable in size with the nuclei of epithelial cells (human fibroblasts) and properly interacts with the surrounding biological environment, without being internalized by cells(**Figure.S1A** and **B**). This would favor the µPL integration with the actual tissue in vivo.

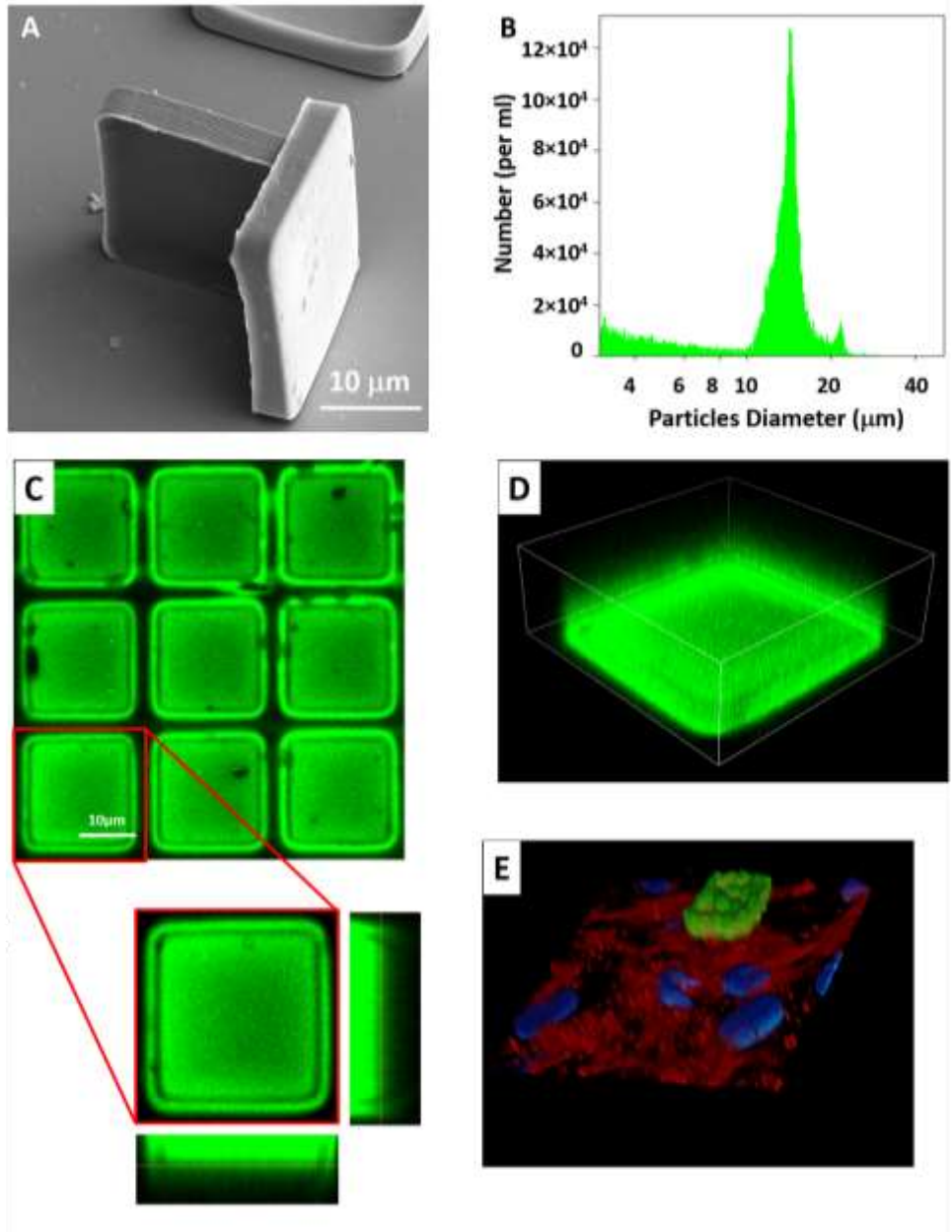

**Figure.2. Geometrical characterization of microplates (µPLs)**. **A**. SEM image of µPLs, showing the characteristic 20 × 20 × 5 µm square shape; **B**. Size characterization of µPLs via Multisizer analysis; **C**. Fluorescence confocal microscopy analysis of µPLs loaded with curcumin (green) with side views; **D**. Fluorescence confocal tri-dimensional reconstruction of a single µPL; **E** CURC-µPLs on a human fibroblast monolayer.

**Biopharmaceutical characterization of Dexamethasone-loaded MicroPlates.** DEX was used as a 'model drug' in that it is one of the most commonly prescribed corticosteroid for the treatment of moderate to severe pains, fever and inflammation, which are symptoms quite common to many pathological conditions.[33] The three different configurations, namely 1, 5 and 7.5 mg PLGA, were characterized in terms of yielding, loading (LE) (**Figure.3A** and **B**), DEX amounts per µPL, and encapsulation efficiency (EE%) (**Figure.S2A** and **B**). For the yielding, which is defined as the percentage of released µPLs over the number of wells in the original silicon template, a similar number of µPL/template was generated in all the three configurations. This was about $6.5 \times 10^5$ particles with a total yielding close to 40% for all considered amounts of PLGA (**Figure.3A**). On the other hand, loading decreases steadily moving from 1.0 to 7.5 mg PLGA µPLs. Specifically, LE is equal to 12.70 ± 1.25% for the 1 mg PLGA particles and decreases up to 4.45 ± 0.70% for the 7.5 mg PLGA particles (**Figure.3B**), in a statistically significant manner ($p < 0.05$). This behavior was expected in that LE is calculated as the weight of the drug over the total weight of the particle, which indeed grows linearly with the amount of PLGA. Furthermore, both the amount of DEX per µPL and the encapsulation efficiency (**Figure.S2A** and **B**) showed a modest advantage for the 5 mg PLGA formulation as compared to the others. Specifically, the DEX loaded amount was 197.38 ± 9.07 pg with an encapsulation efficiency of 11.60 ± 1.45%. This is probably due to the slightly smaller yielding for the 1 mg µPLs together with the denser matrix for the 7.5 mg of PLGA.

The DEX release kinetics from µPLs made out of 1.0, 5.0 and 7.5 mg of PLGA were determined under physiological (PBS; pH 7.4; 37 °C) and pathophysiological (PBS; pH 5.5; 37 °C) conditions, with the latter simulating the acidic microenvironment, typical of an inflamed tissue.[34] Data are shown in **Figure.3C** and **D**, whit grey, orange and blue lines representing the 1.0, 5.0 and 7.5 mg cases, respectively. For the normal condition, about 60% of DEX was released within the first 8 h, whereas the remaining amount of DEX is slowly and continuously released up to 10 days. The initial rapid release should be associated with the drug molecules residing close to the particle surface, whereas the following slow-release phase could rely on the diffusion of the inner molecules towards the outer µPL edges. The drug release profiles for the 1.0, 5.0 and 7.5 mg of PLGA µPLs were similar. The release profiles under pathophysiological conditions (pH 5.5) are given in **Figure.3D**, documenting a slow linear phase followed by an initial burst. The faster release within the first 4 h, as compared to physiological conditions, might depend on the more rapid dissociation of the adsorbed drug, because of its electrostatic nature. These results would suggest that overall the release profile is not strongly influenced by the pH, and thus by polymer degradation within the considered period, but it is mainly controlled by DEX diffusion out of the matrix. In fact, it is well known that PLGA particle degradation is an autocatalytic process which starts from the inside of the matrix.[35] In our experiments, the presence of an acidic microenvironment could mainly promote the surface erosion of the particles and facilitate the escape of the outer DEX molecules.

In order to verify this matrix biodegradation, µPLs loaded with the red fluorescent dye Rhodamine B (RhB-µPLs) were synthetized and characterized over time for possible morphological changes. **Figure.3E** reports bright field, fluorescent and scanning electron microscopy images, taken at different time points, upon µPLs exposure to physiological and pathophysiological conditions. At day 1, RhB-µPLs appeared as square blocks in both cases, but starting from day 3 in acidic conditions, µPL merges appear less straight and, at day 10, µPLs structure appeared roundish and surrounded by spherical microparticles, possibly deriving from the self-assembly process of the eroded polymeric mass. All together these data prove that the degradation process of the particles became a relevant phenomenon only at longer time point, when a significant portion of the drug has been already released.

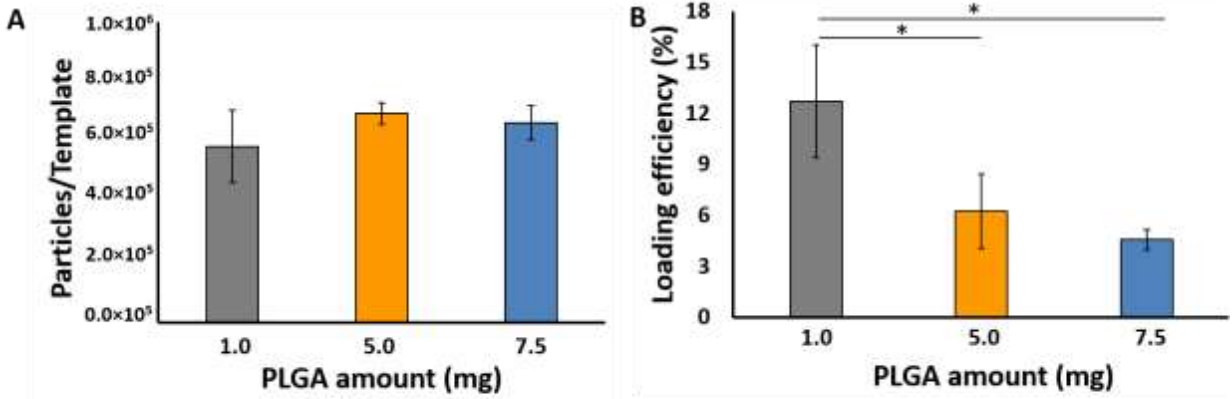
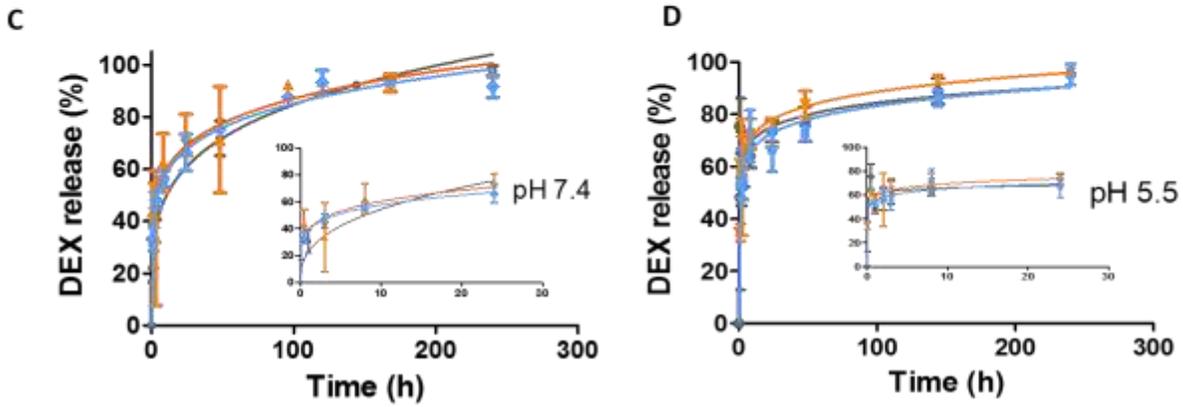
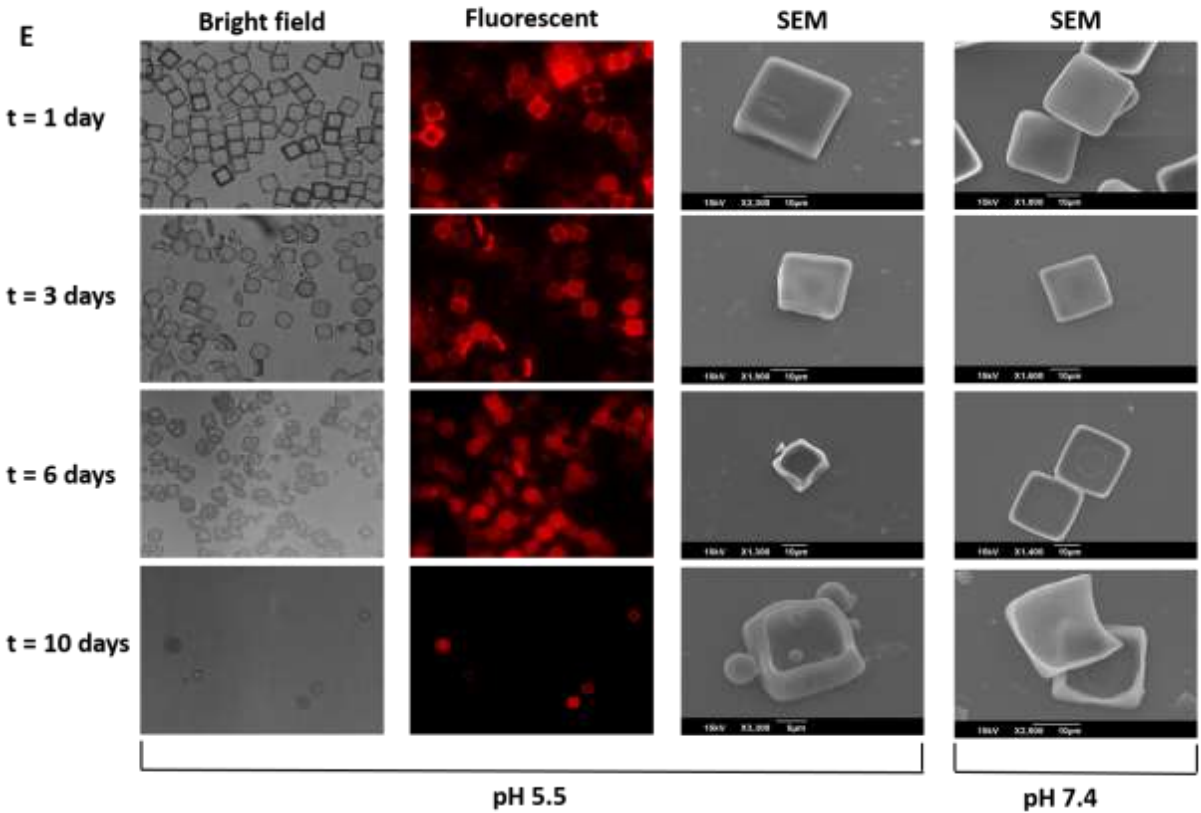

**Figure.3. µPL loading, release and biodegradation. A**. Number of particles yielded from one sacrificial PVA template; **B**. Dexamethasone (DEX) loading into µPLs, using different PLGA amounts; **C**. Release profiles of DEX from µPLs, realized with different PLGA amounts: 1 mg (grey), 5 mg (orange) and 7.5 mg (blue), in PBS buffer at pH 7.4; **D**. Release profiles of DEX from µPLs, realized with different PLGA amount, 1 mg (grey), 5 mg (orange) and 7.5 mg (blue), in PBS buffer at pH 5.5; **E**. degradation of µPLs under pathological (first 3 columns) and physiological (last column) conditions, monitored over time (1, 3, 6, and 10 days) via bright field, fluorescent and scanning electron microscopy. Results are expressed as average ± SD (n = 5). ANOVA analysis: *$p \leq 0.05$.

**Therapeutic efficacy of Dexamethasone-loaded MicroPlates**. µPLs made out of 5 mg of PLGA were selected for all *in vitro* studies, for they have better encapsulation efficiency and higher DEX amount per µPL as compared to the other two configurations. Before studying the interaction of µPLs with inflamed macrophages, the potential toxicity of empty µPLs (**Figure.4A**), free DEX (**Figure.4B**) and DEX-µPLs (**Figure.4C**) was assessed on phagocytic cell lines, such as RAW 264.7 macrophages. Empty particles were tested at a cell to µPL ratios of 1:0.01, 1:0.05, 1:0.1, 1:1 and 1:4. Cell viability was not significantly affected at any of the considered concentrations. Similarly, RAW 264.7 viability was not altered upon 24h incubation with free DEX or DEX-µPLs, even at the greatest tested concentration (30 µM). Note that 30 µM DEX is a concentration 3 times greater than the greatest concentration of DEX used for anti-inflammatory experiments. Also, the incubation time (24 h) was almost 3 times longer than the efficacy experiments, confirming that the DEX-µPLs can be efficiently used as a local drug delivery system. Based on these preliminary considerations, RAW 264.7 and primary bone marrow-derived macrophages (BMDMs) were incubated with DEX-µPL, at 1 µM and 10 µM DEX concentrations, and then stimulated with LPS to induce the secretion of TNF-α, IL-1β and IL-6. Data are presented in **Figure.4D** for RAW 264.7 cells and **Figure.4E** for BMDMs. Data clearly show that DEX-µPLs drastically reduce the secretion of all three considered inflammatory cytokines, in a concentration dependent manner. However, 1 µM of DEX-µPLs is already sufficient to decrease the expression of IL-1β and IL-6 by two and three orders of magnitude, respectively, in the case of BMDMs. Interestingly, the strongest inhibitory effect was registered for the primary macrophages (BMDMs) that appear also to be more extensively stimulated by LPS as compared to RAW 264.7 cells.

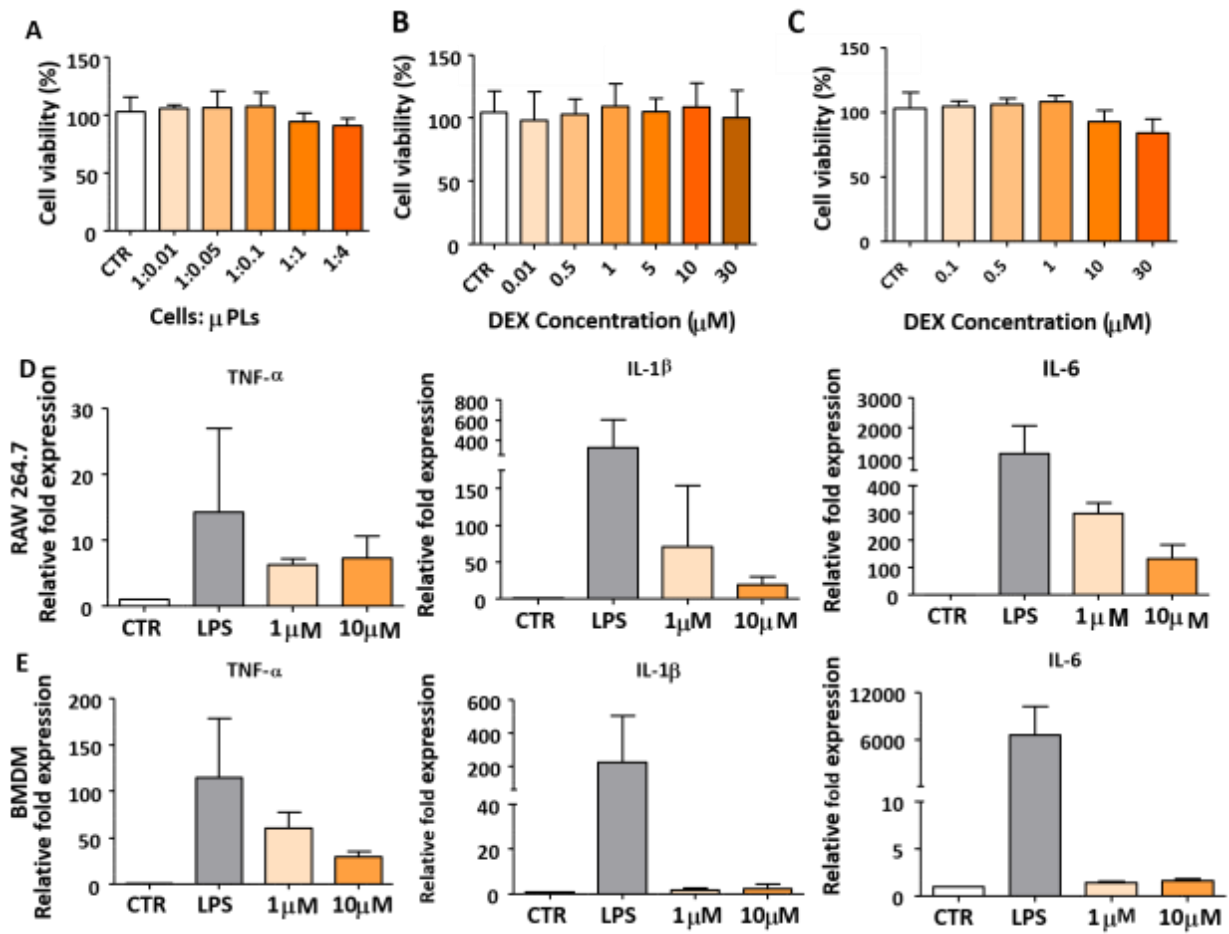

**Figure.4. Anti-inflammatory efficacy of DEX-µPLs. A-C**. Viability of RAW 267.4 cells incubated with empty µPLs, free DEX and DEX-µPLs, respectively. **D**. and **E**. Pro-inflammatory cytokines expression levels for LPS stimulated RAW 267.4 and BMDM cells, respectively. (CTR: no LPS and no µPLs). Results are expressed as average ± SD (n = 3).

**Mechanical characterization of MicroPlates.** µPLs are large, micron-sized particles that cannot be administered systemically but would rather function as local depots for the continuous and controlled release of therapeutic molecules and possibly nanoparticles. As such, the ability to modulate µPLs stiffness is a relevant factor in that it would facilitate their integration with the surrounding environment. Following this notion, the PLGA amount in the polymeric mixture for the µPLs preparation was changed, returning particles with different mechanical properties (**Figure.5**). First, atomic force microscopy (AFM) was used to assess µPLs morphology (**Figure.5D**), confirming overall the geometrical features already observed via scanning electron and fluorescent microscopies. Then, via AFM quantitative imaging mode, multiple force-displacement curves were generated by indenting the µPLs at different spots. **Figure.5E** shows a representative force-displacement curve. As reported in **Figure.5F**, µPL stiffness increased with an increasing amount of PLGA, namely from 0.8 MPa for 1 mg of PLGA to 5.7 MPa for 7.5 mg of PLGA. These values fall well within the Young's modulus of native tissues.[36] For instance, the stiffness value for 5 mg PLGA µPLs, which is of about 2.08 ± 0.5 MPa, is similar to the one of cartilage tissues.[37] Nevertheless, this modulation in mechanical properties, by changing PLGA amount, do not affect µPLs size and shape. As from SEM images, besides a very moderate increase in central concavity, there is no morphological difference among µPLs realized with 1, 5 and 7.5 mg of PLGA (**Figure.5A-C**, respectively).

The observed trend for the stiffness-PLGA amount relationship is in agreement with the well accepted notion that hydrogel stiffness can be modulated by changing polymer concentration.[38] Noteworthy, for polymeric nanoparticles obtained via a conventional bottom-up approach, an increase in polymer mass is generally associated with an increase in hydrodynamic diameter.[39] In other words, the proposed top-down approach allows us to control the geometry and the mechanical properties of the particles simultaneously and independently.

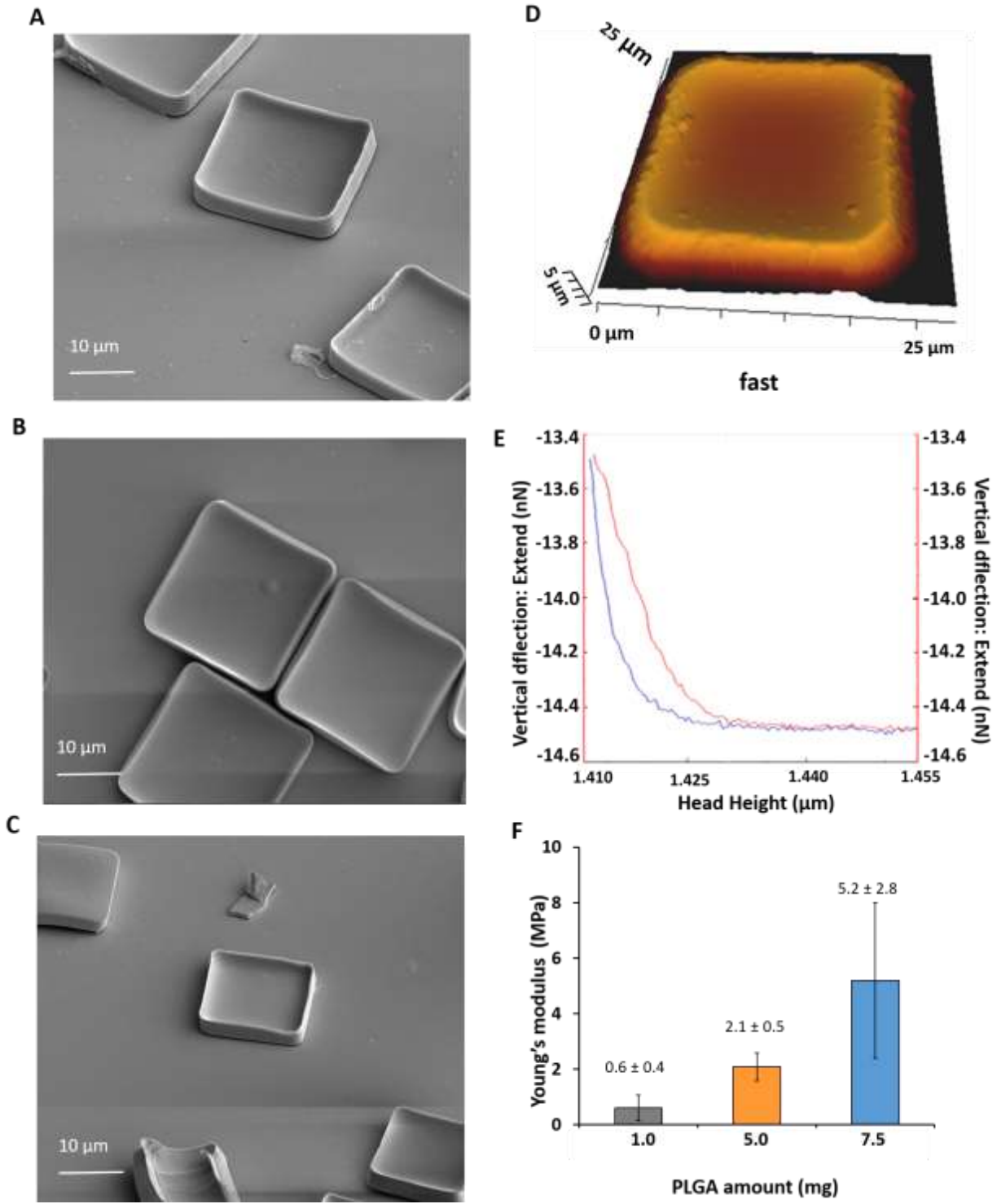

**Figure.5 Mechanical characterization of µPLs. A-C**. SEM images of µPLs made out of 1.0, 5.0 and 7.5 mg of PLGA, respectively **D**. Morphology of a µPL obtained via atomic force microscopy

(AFM) analysis. **E**. Representative force–displacement curve. Red curve represents the extension, while the blue one the retraction of AFM cantilever. **F**. Young's modulus of µPLs realized with different amount of PLGA, namely 1.0, 5.0 and 7.5 mg.

**Preparation and characterization of DEX-SPNs-µPLs.** Being conceived as a local drug delivery system, the size of µPLs is not appropriate for parenteral administration. However, multiple and different nanoparticles can be distributed, together with free drug molecules, within the µPL matrix, giving rise to a hierarchical multiscale system. As an example, drug-loaded spherical polymeric nanoconstructs (SPNs) were distributed within µPLs, thus realizing a hierarchical nanoconstruct spanning from the molecular size of the therapeutic payload to the nano-sized SPNs and micron-sized µPLs (**Figure.6A**). DEX-SPNs present an average size of $199.98 \pm 25.87$, a PDI value of $0.080 \pm 0.010$, and a $\zeta$-potential of $-40.50 \pm 5.50$ mV. DEX-SPNs were also characterized in terms of loading, returning a LE = $3.47 \pm 1.22\%$ and EE = $5.42 \pm 0.74\%$.

DEX-SPNs were dispersed within the polymeric paste used for assembling the µPLs and deposited over the template following the same approach used for DEX-µPLs. However, prior to their dispersion in the paste, DEX-SPNs were protected by a thin PVA coating which did not alter their physicochemical properties (see **Figure.S3**). At this point, the release of dexamethasone acetate from DEX-SPNs, DEX-µPLs, and DEX-SPNs-µPLs was studied. A direct comparison is shown in **Figure.6B**. For DEX-SPNs, about 90% of DEX was released within the first 8 h, whereas the remaining 10% was released within the following 48h. In the case of DEX-µPLs, only 60% of DEX was released during the first 8h, followed by a slow release until 10 days. Finally, the release from DEX-SPNs-µPLs was much slower: at 8h, only around 35% of DEX was released. This represents about 0.5 and 0.25 times the drug amounts released from DEX-SPNs and DEX-µPLs, respectively. This trend slower release is observed throughout the whole experiment (240 h). Based on fitting considerations, the three release data are statistically different ($p$ value < 0.0001). The entrapment efficiency of DEX-SPNs inside µPLs slows down DEX release, probably for the presence of the polymer matrix surrounding DEX-SPNs. In this case, water has to first penetrate and hydrate the µPLs matrix, and, after that, reach and hydrate the SPNs, as to free the drug molecules and starts their diffusion within the two polymeric matrices. Furthermore, the presence of the PVA surfactant on SPNs could contribute to this effect. The use of SPNs as drug carriers and loading them inside the µPLs represents a strategy for modulating the release, especially within the first h, and for realizing truly multifunctional systems.

To document the encapsulation and stability of SPNs within the µPL matrix, confocal fluorescent microscopy and transmission electron images were generated for SPNs loaded with gold nanoparticles and labeled with RhB (**Figure.6C** and **D**, and **Figure.S4)**. The 20 nm spherical gold nanoparticles are clearly visible within the µPL matrix, whose particular structure can be appreciated in the TEM image of **Figure.6C**. The confocal image shows distinct red spots, rather than a diffuse fluorescence, confirming the intact structure and homogeneous distribution of SPNs after loading in the µPLs (**Figure.6 D** and **Figure.S4)**.

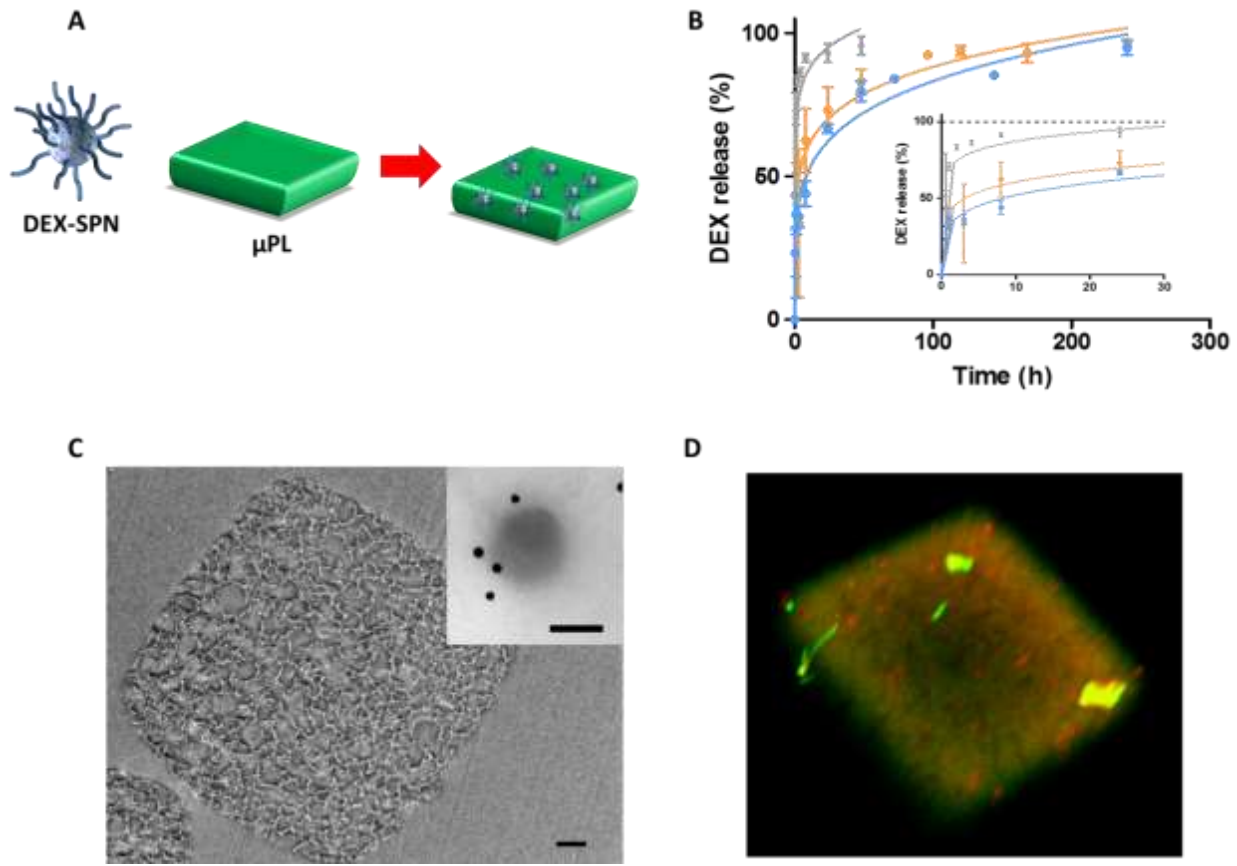

**Figure.6. Hierarchical, multifunctional µPLs. A**. Schematic representation of µPL, DEX-SPN and DEX- SPNs-µPL. **B**. Release profile of DEX from DEX-SPNs (grey), DEX-µPLs (orange) and DEX-SPNs-µPLs (blue). Results are expressed as average ± SD (n = 3). **C**. TEM analysis of Au-Cy5-SPNs µPLs (scale bar 2µm). In the inset black spheres represent Au nanoparticles (200 nm). **D**. Representative confocal tridimensional reconstruction of Au-Cy5-SPNs loaded inside one µPLs, showing their homogeneous distribution.

## Conclusion

Micron-sized polymeric particles – the µPLs – were realized using a top-down fabrication approach that allows us to control precisely and independently their geometrical, mechanical and pharmacological properties. μPLs have a square base with an edge length of 20 μm and a height of 5 μm. In previous works, using the same fabrication strategy, the authors have shown that particle geometry can be readily tailored.[26-27] Furthermore, controlling the amount of PLGA allow us to realize μPLs with different mechanical stiffness, without affecting the actual particle geometry. With a PLGA amount ranging from 1 to 7.5 mg, the μPL Young's modulus changed by one order of magnitude, namely from 0.6 MPa up to 5 MPa. The particle deformability is crucial in facilitating their deposition and integration with the surrounding tissues.

In terms of pharmacological properties, the anti-inflammatory molecule Dexamethasone was dispersed within the μPL matrix, and sustained release over a period of 10 days was demonstrated, with a significant burst within the first few h followed by a steady, continuous release. It was also shown that μPL could be loaded with nanoparticles whereby such a hierarchical system can be used for building, in a modular way, multiple functionalities in the μPLs. Specifically, it was demonstrated that the release profile of DEX could be modulated by dispersing within the μPL matrix spherical polymeric nanoconstructs, loaded themselves with DEX, thus diminishing the initial burst release and modulating release throughout the whole process.

Our findings evidence that μPLs can be used as a drug depot, combining sustained release profile, needed for all local drug delivery systems, with a precise control in geometrical and mechanical properties, fostering optimal tissue implantation.

**ACKNOWLEDGEMENTS**

This project was partially supported by the European Research Council (FP7/2007-2013)/ERC - grant 616695); the European Commission under the Marie Skłodowska-Curie actions (H2020-MSCA-COFUND-2016 - grant 754490) and the Italian Association for Cancer Research (individual investigator - grant 17664). The authors acknowledge the precious support provided by the Nikon Center and the Electron Microscopy Facility at the Italian Institute of Technology for microscopy acquisitions and analyses.

# Supporting Information

## HIERARCHICAL MICROPLATES AS DRUG DEPOTS WITH CONTROLLED GEOMETRY, RIGIDITY AND THERAPEUTIC EFFICACY

Martina Di Francesco[1], Rosita Primavera[1], Davide Romanelli[1], Roberto Palomba[1], Tiziano Catelani[1], Christian Celia[2], Luisa Di Marzio[2], Massimo Fresta[3], Daniele Di Mascolo[1], Paolo Decuzzi[1],♣

[1] Laboratory of Nanotechnology for Precision Medicine,
Fondazione Istituto Italiano di Tecnologia,
Via Morego 30, Genoa 16163, Italy

[2] Department of Pharmacy, University of Chieti – Pescara "G. D'Annunzio",
Via dei Vestini - Campus universitario - 66100 Chieti, Italy

[3] Department of Health Sciences,
University of Catanzaro "Magna Graecia", Viale Europa - 88100 Catanzaro, Italy

♣ To whom correspondence should be addressed: Paolo Decuzzi, PhD. Phone: +39 010 71781 941, Fax: +39 010 71781 228, E-mail: paolo.decuzzi@iit.it

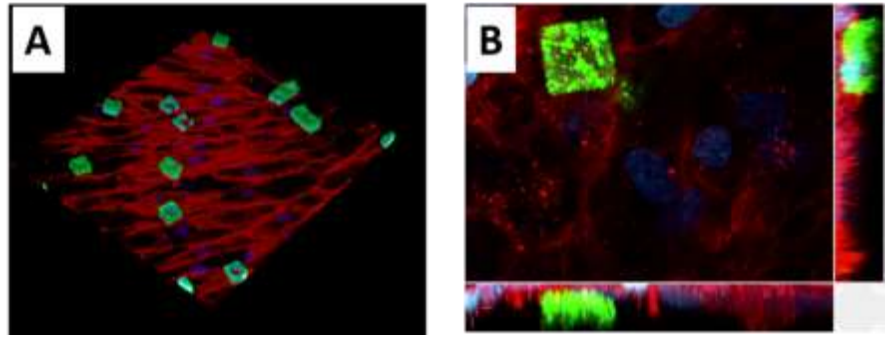

**Figure.S1. µPLs interaction with human fibroblasts monolayer. A**. 3D confocal microscopy analysis of human fibroblasts cell after incubation with CURC-µPLs; **B**. representative Z-stack analysis demonstrating that CURC-µPLs are not internalized by human fibroblast but rather lodge on the cell membrane.

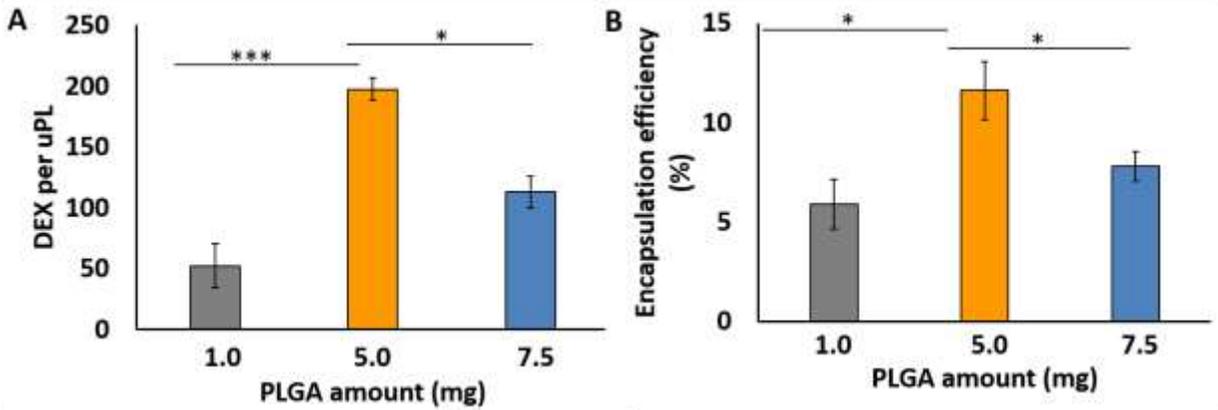

**Figure.S2. Effect of PLGA amounts on µPLs loading. A.** Mass of DEX loaded per µPL; **B**. Entrapment efficiency of DEX into µPLs. Results are expressed as average ± SD (n = 5). ANOVA analysis: *p ≤0.05, ***p ≤0.001

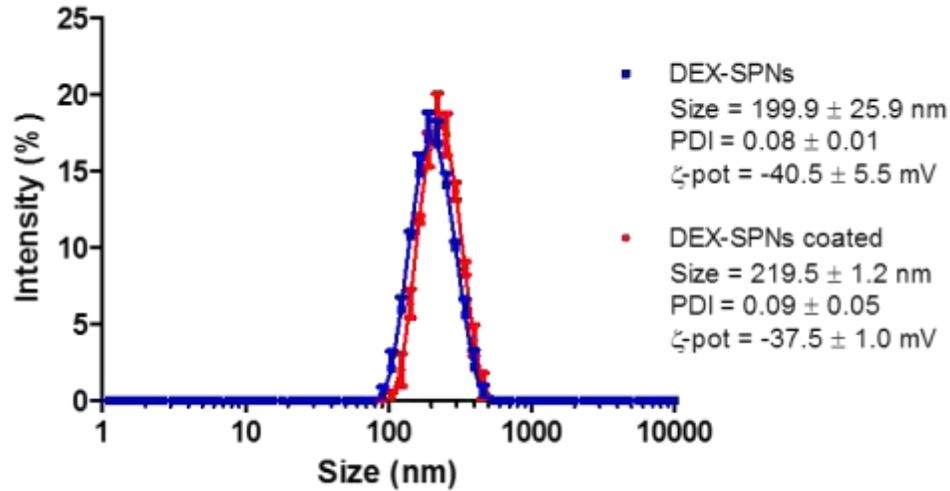

**Figure.S3. Effect of PVA coating on DEX-SPN physicochemical properties.** DLS size distribution of DEX-SPNs (blue line) and PVA-coated DEX-SPNs (red line), showing a very modest increase in size, due to the surface PVA coating. This is also confirmed by the slight increase in ζ-potential.

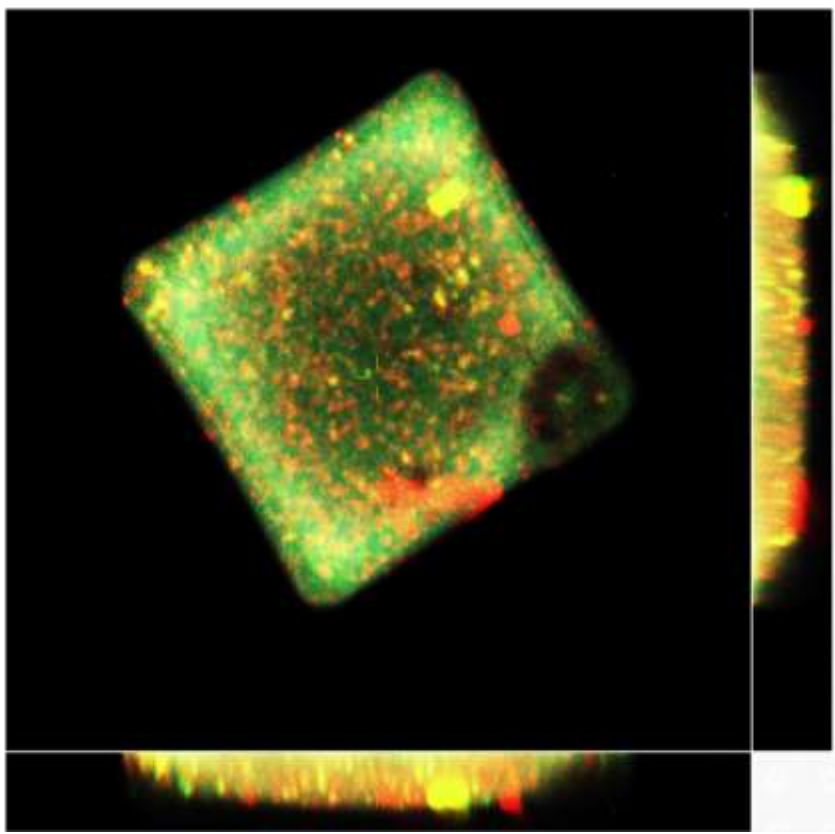

**Figure.S4. Distribution of SPNs within the μPL matrix.** Representative confocal Z-stack image of a μPL loaded with molecular curcumin (diffuse green color) and RhB-labeled SPNs (red spots). The latter are uniformly distributed within the green matrix of μPLs